\documentclass[floatfix,letterpaper,english,preprint, secnumarabic,amssymb,nobibnotes, aps, prd,superscriptaddress]{revtex4-2}
\usepackage[T1]{fontenc}
\usepackage[latin9]{inputenc}
\usepackage{babel}
\usepackage{amsmath}
\usepackage{amssymb}
\usepackage{cancel}
\usepackage{graphicx}
\usepackage{wasysym}
\usepackage{physics}
\usepackage{multirow}
\usepackage{soul}
\usepackage[unicode=true,pdfusetitle,bookmarks=true,bookmarksnumbered=false,bookmarksopen=false,breaklinks=false,pdfborder={0 0 1},backref=false,colorlinks,linkcolor=blue,anchorcolor=blue,citecolor=blue]{hyperref}

\makeatletter

\pdfpageheight\paperheight
\pdfpagewidth\paperwidth

\makeatother

\begin{document}
\title{Fast Fermion Smearing Scheme with Gaussian-like Profile}

\author{Chuan-Yang Li}
\affiliation{Key Laboratory of Atomic and Subatomic Structure and Quantum Control (MOE), Guangdong Basic Research Center of Excellence for Structure and Fundamental Interactions of Matter, Institute of Quantum Matter, South China Normal University, Guangzhou 510006, China}
\affiliation{Guangdong-Hong Kong Joint Laboratory of Quantum Matter, Guangdong Provincial Key Laboratory of Nuclear Science, Southern Nuclear Science Computing Center, South China Normal University, Guangzhou 510006, China}

\author{Terrence Draper}
\affiliation{Department of Physics and Astronomy, University of Kentucky, Lexington,
KY 40506, USA}

\author{Jun Hua}
\affiliation{Key Laboratory of Atomic and Subatomic Structure and Quantum Control (MOE), Guangdong Basic Research Center of Excellence for Structure and Fundamental Interactions of Matter, Institute of Quantum Matter, South China Normal University, Guangzhou 510006, China}
\affiliation{Guangdong-Hong Kong Joint Laboratory of Quantum Matter, Guangdong Provincial Key Laboratory of Nuclear Science, Southern Nuclear Science Computing Center, South China Normal University, Guangzhou 510006, China}

\author{Jian Liang}
\email{jianliang@scnu.edu.cn}
\affiliation{Key Laboratory of Atomic and Subatomic Structure and Quantum Control (MOE), Guangdong Basic Research Center of Excellence for Structure and Fundamental Interactions of Matter, Institute of Quantum Matter, South China Normal University, Guangzhou 510006, China}
\affiliation{Guangdong-Hong Kong Joint Laboratory of Quantum Matter, Guangdong Provincial Key Laboratory of Nuclear Science, Southern Nuclear Science Computing Center, South China Normal University, Guangzhou 510006, China}

\author{Keh-Fei~Liu}
\affiliation{Department of Physics and Astronomy, University of Kentucky, Lexington,
KY 40506, USA}

\author{Jun Shi}
\email{jun.shi@scnu.edu.cn}
\affiliation{Key Laboratory of Atomic and Subatomic Structure and Quantum Control (MOE), Guangdong Basic Research Center of Excellence for Structure and Fundamental Interactions of Matter, Institute of Quantum Matter, South China Normal University, Guangzhou 510006, China}
\affiliation{Guangdong-Hong Kong Joint Laboratory of Quantum Matter, Guangdong Provincial Key Laboratory of Nuclear Science, Southern Nuclear Science Computing Center, South China Normal University, Guangzhou 510006, China}

\author{Nan Wang}
\affiliation{Key Laboratory of Atomic and Subatomic Structure and Quantum Control (MOE), Guangdong Basic Research Center of Excellence for Structure and Fundamental Interactions of Matter, Institute of Quantum Matter, South China Normal University, Guangzhou 510006, China}
\affiliation{Guangdong-Hong Kong Joint Laboratory of Quantum Matter, Guangdong Provincial Key Laboratory of Nuclear Science, Southern Nuclear Science Computing Center, South China Normal University, Guangzhou 510006, China}

\author{Yi-bo Yang}
\email{ybyang@itp.ac.cn}
\affiliation{CAS Key Laboratory of Theoretical Physics, Institute of Theoretical Physics, Chinese Academy of Sciences, Beijing 100190, China}
\affiliation{School of Fundamental Physics and Mathematical Sciences, Hangzhou Institute for Advanced Study, UCAS, Hangzhou 310024, China}
\affiliation{International Centre for Theoretical Physics Asia-Pacific, Beijing/Hangzhou, China}
\affiliation{University of Chinese Academy of Sciences, School of Physical Sciences, Beijing 100049, China}

\begin{abstract}
We propose a novel smearing scheme which
gives a Gaussian-like profile and is more efficient than the traditional Gaussian smearing
in terms of computer time consumption.
We also carry out a detailed analysis of the profiles, smearing sizes, and the behaviors of hadron effective masses of different smearing schemes, and point out that having a sufficient number of gauge paths in a smearing scheme is essential to produce strong smearing effects.
For practical smearing sizes $\bar{r}= 10a \sim 20a$, the novel smearing reduces the time cost 
by about $8 \sim 10$ times compared with the traditional Gaussian smearing.
This path smearing scheme and
its variants would be beneficial for lattice studies of hadron
spectra and structures.

\end{abstract}

\maketitle

\section{Introduction}

Lattice QCD simulation provides a systematic approach to study the strong interaction non-perturbatively,
which has achieved remarkable success in the past few decades.
The basic quantities calculable on the lattice are
$n$-point correlation functions in the temporal direction,
and the widely-seen exponential behavior
of the correlation functions is rooted in the 
Euclidean signature of the theory.
Usually, physical observables, like hadron energies and hadron matrix elements,
are extracted from the correlation functions by fitting to some exponential forms.
However, the rapidly decreasing signal-to-noise ratio (S/N) with increasing Euclidean time separations
and the fact that the correlation functions contain mixed contributions from all the hadronic states of the same 
quantum number pose the main obstacles in the numerical analyses of lattice correlators.
Currently, with regard to the ground state physics, 
the most effective prescription for these underlying difficulties
is to smear the fermion fields in the hadronic interpolating operators.
Smearing suppresses the excited-state contribution and enables the fit windows to include earlier time slices
with better S/N.
Another example is
the so-called momentum smearing~\cite{Bali:2016lva},
which helps to increase significantly the S/N for correlators of highly boosted hadrons
and has become one of the most important techniques in lattice studies of parton physics.

Most commonly used smearing schemes, such as the Jacobi smearing~\cite{Collins:1992fj,UKQCD:1993gym} and
Wuppertal smearing~\cite{Gusken:1989qx}, often lead to a Gaussian profile.
Several smearing schemes resulting in different profile shapes
have also been discussed in the literature, e.g.,  Refs.~\cite{Bacilieri:1988fh,Allton:1990qg,DeGrand:1991ng,vonHippel:2013yfa}.
To observe the requirement of gauge invariance of the hadronic operators, 
the kernel of these smearing schemes is 
implemented gauge covariantly in an iterative manner.
According to some numerical checks~\cite{XQCD:2013odc},
the number of iteration needed
grows non-linearly as 
the size of smearing increases.
And for a particular physical smearing size,
smaller lattice spacings demand more 
iterations.
Thus performing a large fermion smearing on
fine lattices is in some sense costly.
Certainly, the inversion
of the fermion matrix is always
the most expensive part in lattice calculations.
But at the same time,
much effort has been focused on
the corresponding optimization.
Nowadays, with the help of advanced
software and hardware, the inversion operation is
accelerated by hundreds of times, 
but the smearing algorithm has relatively lacked attention.
In some practical calculations,
the cost of smearing becomes compatible with that of computing the propagators.
Especially for the calculations using overlap fermions, e.g. Refs.~\cite{Yang:2018nqn,Liang:2023jfj}, with low-mode substitution technique~\cite{xQCD:2010pnl},
the smearing needs to be done many times for each eigenvector
which is one of the main sources of time cost.

One idea of solving this problem
which should be mentioned here 
is the smearing method by using
Laplace eigenvectors~\cite{HadronSpectrum:2009krc,Morningstar:2011ka}. However, although
this method is widely used in the lattice studies of spectroscopy,
it is not directly feasible for the three-point function calculations
with a local current operator.
We have also proposed and used a fast fermion smearing algorithm~\cite{Liang:2016fgy,Liang:2018pis},
which gives a quasi block-shaped profile and is called the naive block smearing in this article.
This smearing scheme can generate efficiently a large smearing size as
other smearing schemes, however, in practical hadron two-point correlation function
calculations,
it has problems in producing strong smearing effects.
This is attributed to the limited number of gauge paths connecting different
lattice sizes in this algorithm which will be discussed in detail in Sec.~\ref{sec:novel_smearing}.

In this article,
based on the naive block smearing and 
the understanding on the role of number of gauge paths in the smearing kernel,
we propose a novel gauge covariant fermion smearing scheme.
The profile of this new smearing is Gaussian-like, which
is different from that of the naive block smearing.
It keeps the efficiency of the naive block smearing, while overcoming the difficulties in
generating strong smearing effects.
The improvement is remarkable especially for large smearing on fine lattices.
And it can be used potentially to replace
the traditional smearing schemes in lattice studies to
meliorate the calculation procedure.
The numerical calculations in this study are done on two $2+1$-flavor gauge ensembles of
clover fermions generated by the CLQCD collaboration~\cite{Hu:2023jet}. 
One is a small $24^3\times 72$ lattice with pion mass $\sim$ 290 MeV and lattice spacing $\sim$ 0.11 fm, and the other is
a finer lattice of size $48^3\times 144$ with pion mass $\sim$ 317 MeV and lattice spacing $\sim$ 0.05 fm.
Most calculations are done on the first lattice while
the finer lattice is used as a complement to further demonstrate the efficiency of the novel smearing scheme.
One-step hypercubic (HYP)~\cite{Hasenfratz:2001hp} smeared gauge links are used in all the fermion smearings.

This article is organized as follows.
First in Sec.~\ref{sec:naive_block}, a detailed description of the algorithm for the naive block smearing is presented.
We then carry out a detailed comparison between the naive block smearing and the traditional Gaussian smearing on the profile, smearing size, and the behavior of hadron effective masses.
An analysis of the reason of the limitation of the naive block smearing is also performed.
In Sec.~\ref{sec:novel_smearing}, based on the naive block smearing, we construct a novel smearing scheme and demonstrate that it is effective and efficient both formally and numerically.
We point out that having a sufficient number of gauge paths is essential for a smearing scheme to produce strong smearing effects.
Finally, Sec.~\ref{sec:summary} gives a summary and discussion of this study.

\section{the Naive Block Smearing} 
\label{sec:naive_block}

Physically, the block-shaped smearing scheme
can be treated as a variant of the so-called wall source,
which has a constant value on each spatial point before momentum projection.
It is well-known that, when using gauge-fixed wall sources, 
the effective mass of hadron two-point functions (wall source to point sink) approaches to the ground-state energy from below as the Euclidean time increases,
which is due to negative overlapping factors of the excited states,
and can be interpreted by the hadronic Bethe-Salpeter wave functions phenomenologically. 
When focusing on the ground-state physics,
one would like to have near-zero excited-state contamination rather than the negative contribution.
A direct idea therefore is to reduce the size of the ``wall'',
which results in a spatial block in the source time slice.
Technically, 
a fast gauge-covariant block smearing algorithm
has been developed~\cite{Liang:2016fgy}
to speed up the smearing processes in practical calculations.
Since this algorithm is the core of the naive block smearing, 
and is essential to understand the implementation of the novel smearing scheme,
we elaborate its idea and procedure as follows for the readers' convenience.

The key building blocks of smearings are the forward and backward
shift operations ${\cal S}^+_i[\psi(x)]=\psi(x)\to U_i(x)\psi(x+\hat{i})$ and
${\cal S}^-_i[\psi(x)]=\psi(x)\to U_i^\dagger(x-\hat{i})\psi(x-\hat{i})$,
which shift the fermion field $\psi(x)$ on the entire lattice by one lattice spacing and $\hat{i}$ denotes the unit vector along direction $i$.
Since these are the basic routines in modern lattice codes,
the time consumption of a smearing scheme can be 
measured by counting the number of shift operations used.
For the conventional Gaussian smearing,
the smearing operator can be written in terms of the shift operations as
\begin{equation}
    \label{eq:gaussian}
    \hat{G} \sim \left [1-\frac{3\omega^2}{2n} + \frac{\omega^2}{4n} \sum_{i=1}^{3} \left({\cal S}^+_i +{\cal S}^-_i\right) \right] ^n,
\end{equation}
where $\omega$ is the smearing parameter and $n$ indicates the
number of iterations.
Obviously, $6n$ shift operations are needed in this scheme for $n$ times of iteration.
Analogously, the naive block smearing can be expressed in the following form
\begin{widetext}
\begin{equation}
    \label{eq:block}
    \hat{B} \sim \frac{1}{6} \sum_{P(i,j,k)}
    \left [1 + \sum_{m=1}^{n} \left(({\cal S}^+_i)^m +  ({\cal S}^-_i)^m\right) \right]
    \left [1 + \sum_{m=1}^{n} \left(({\cal S}^+_j)^m +  ({\cal S}^-_j)^m\right) \right]
    \left [1 + \sum_{m=1}^{n} \left(({\cal S}^+_k)^m +  ({\cal S}^-_k)^m\right) \right],
\end{equation}
\end{widetext}
where the average over the permutations of the spatial directions $P(i,j,k)$ is
actually to average the six different gauge paths (e.g. $(i,j,k)=(1, 2, 3)$ means the gauge path goes in the $x$ direction first, and then the $y$ direction, and finally $z$),
which is to avoid the potential breaking of rotation symmetry.
Thus the number of shift operations required is $6\times6n$.

\begin{figure*}[h]
    \centering    
    \includegraphics[width=0.85\textwidth]{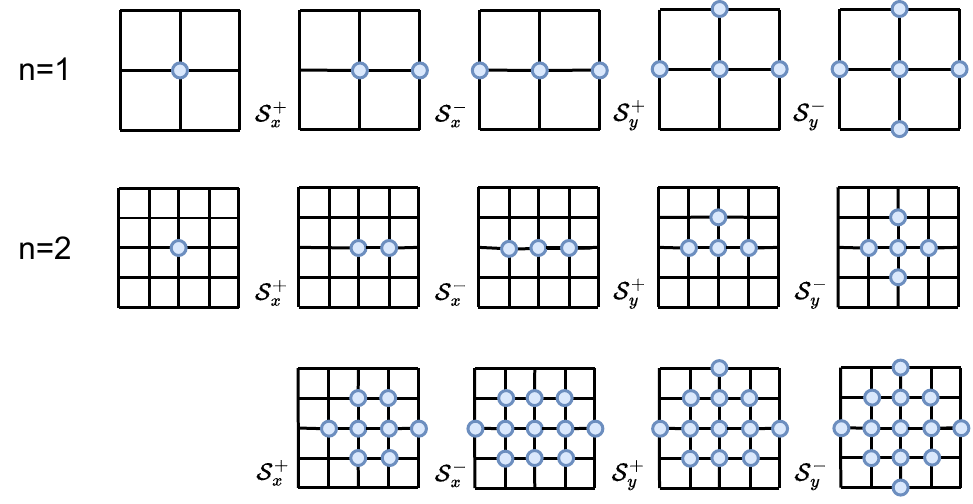}
    \caption{Schematics of the smearing procedures on a 2-D grid for the Gaussian smearing with $n=1$ and $2$ where the blue dots indicate that the field has nonzero values on the corresponding points.}
    \label{fig_gaussian_cartoon}
\end{figure*}

\begin{figure*}[h]
    \centering    
    \includegraphics[width=0.85\textwidth]{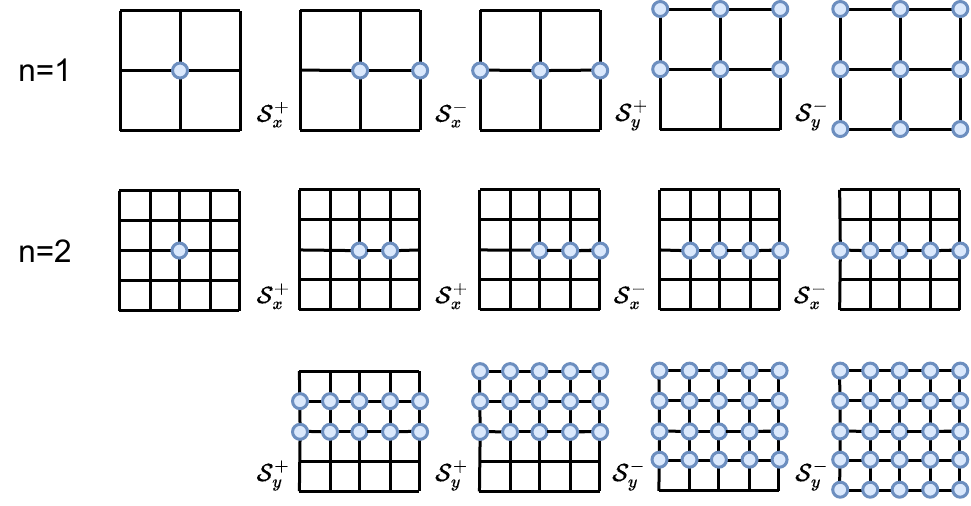}
    \caption{Same as Fig.~\ref{fig_gaussian_cartoon} but for the naive block smearing.}
    \label{fig_block_cartoon}
\end{figure*}

Comparing with the Gaussian smearing, one main distinction of the
naive block smearing is that
the shift operations are done separately in each direction.
Note that, as mentioned before, the shift operations are for fields on the whole lattice, therefore the shifts along the next direction are on top of the extended fields resulted from the previous ones, leading to a fast growth of the smeared range.
The schematic smearing procedures on a 2-D lattice for Gaussian and the naive block smearings when $n=1$ and $2$ are depicted in Fig.~\ref{fig_gaussian_cartoon} and Fig.~\ref{fig_block_cartoon} respectively. We can see that the Gaussian smearing takes one shift along $\pm x$ and $\pm y$ directions as a basic step and repeats this step on top of the previous ones. Whereas the naive block smearing takes all $n$
times of shifts along one axis as a basic step, resulting in a block smearing shape.

\begin{figure}[h]
    \centering
    \includegraphics[width=0.45\textwidth]{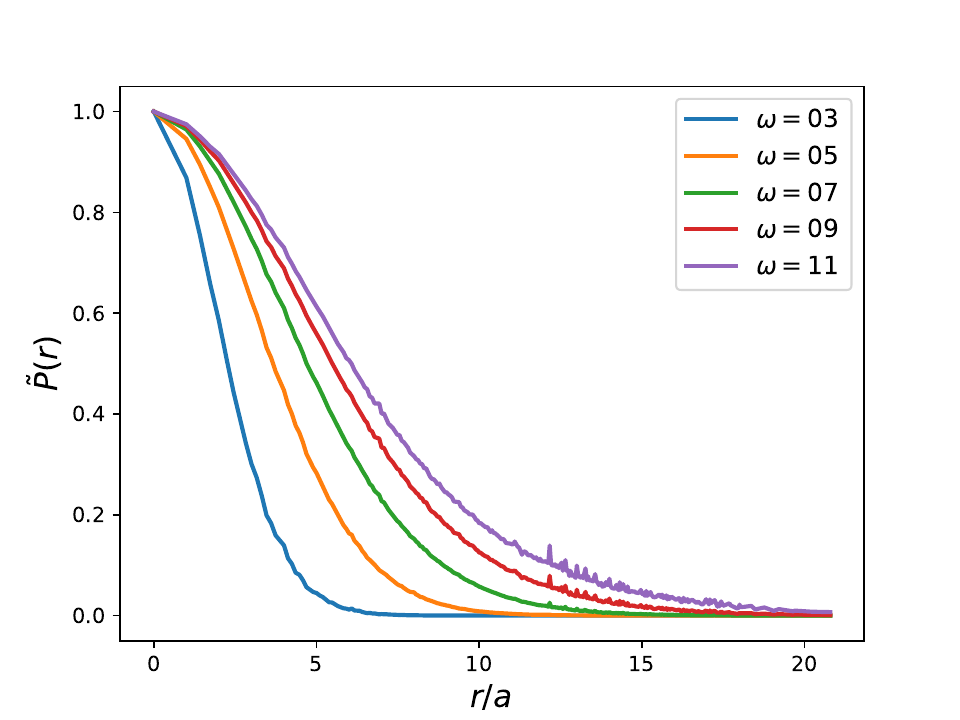}
    \includegraphics[width=0.45\textwidth]{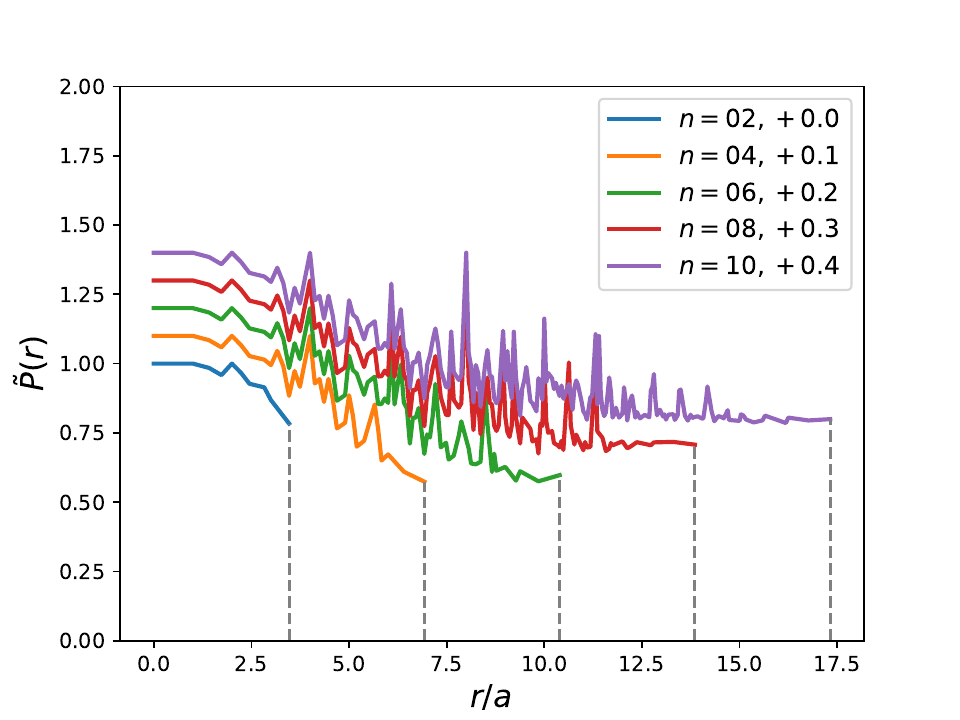}
    \caption{The normalized profiles of the Gaussian (left) and the naive block (right) smearings. The ones of the naive block smearing with different $n$ are shifted vertically for better visualization as indicated by the legends.}
    \label{fig_prfl_gaussian_block}
\end{figure}

To further compare the Gaussian and the naive block smearings, we look at their smearing profiles, which are defined as
\begin{equation}
    {P}(r) = \frac{1}{N_r} \sum_{\abs{\vec{x}-\vec{x}_0} = r} \sqrt{\frac{1}{3}\Tr_c\left[S(\vec{x},\vec{x}_0)S^{\dagger}(\vec{x},\vec{x}_0)\right] }
    \label{eq:P_r}
\end{equation}
where $S(\vec{x},\vec{x}_0)$ is the smearing function representing a smeared quark field from a point source located at $\vec{x}_0$, $N_r$ is a normalization factor which is the number of points satisfying the condition $\abs{\vec{x}-\vec{x}_0} = r$,
and the trace lives in color space (the Dirac indices are omitted since they are irrelevant to smearings). 
We use $S(\vec{x},\vec{x}_0)S^{\dagger}(\vec{x},\vec{x}_0)$ to guarantee the profile is gauge invariant.
The normalized profiles $\tilde{P}(r)\equiv P(r)/P(0)$ of the Gaussian and the naive block smearings are shown in Fig.~\ref{fig_prfl_gaussian_block}.
The figures show that the profile of the Gaussian smearing is indeed Gaussian shaped, whereas that of the naive block smearing is not exactly block shaped but with oscillations and a slightly descending behavior due to the effects of gauge links.
To be specific, for each $\vec{x}$, the smearing function of the naive block smearing contains 6 gauge paths (each path is a production of $SU(3)$ matrices) due to the summation over the permutations in all three spatial directions.
For certain points at $r = 0,~1,~2,~4,~8,~\dots$, all the 6 paths are
degenerate, such that the $S(\vec{x},\vec{x}_0)S^{\dagger}(\vec{x},\vec{x}_0)$ gives a unit $3\times 3$ matrix, and the profile equals one as shown in the right panel of Fig.~\ref{fig_prfl_gaussian_block}. For other points, the smearing function contains a linear combination of different paths, such that $\frac{1}{3}\Tr_c\left[S(\vec{x},\vec{x}_0)S^{\dagger}(\vec{x},\vec{x}_0)\right]<1$. The oscillation reflects the number of different paths at different points. On the other hand, the decaying behavior is due to the fact that the gauge matrices are not identity. A rough fit of form $\tilde{P}(r)\propto A^r$ gives $A\sim 0.93$ which is close to the plaquette value of this lattice.

\begin{figure}[htbp]
    \centering    
    \includegraphics[width=0.45\textwidth]{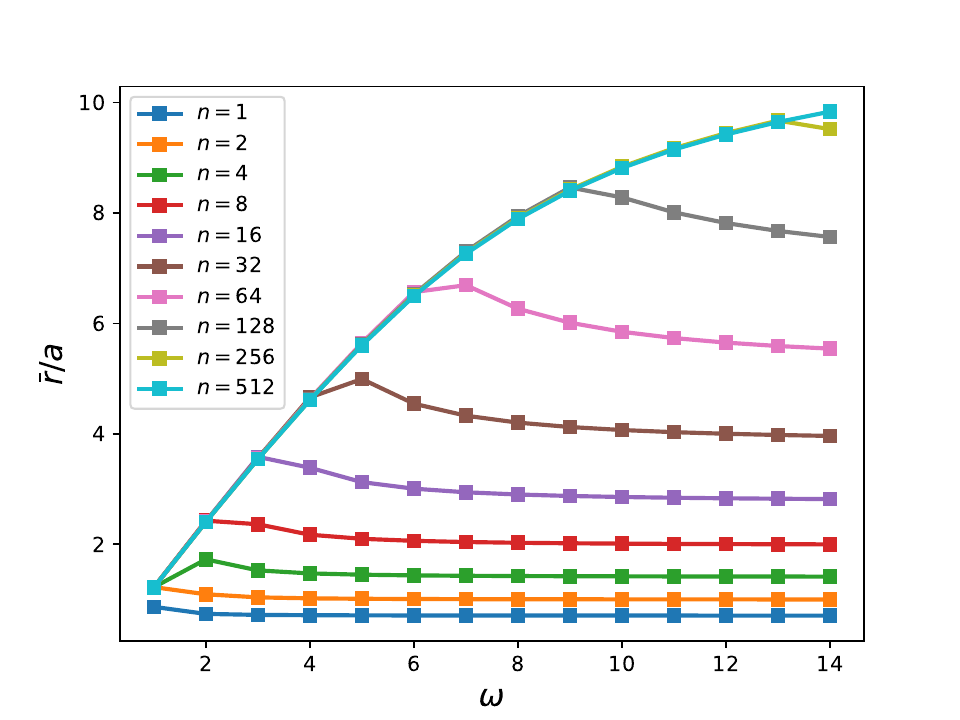}
    \caption{Smearing size $\bar{r}$ at different $\omega$ and iteration time $n$ in Gaussian smearing.}
    \label{fig_gs_w_r_n}
\end{figure}

\begin{figure}[h]
    \centering    
    \includegraphics[width=0.45\textwidth]{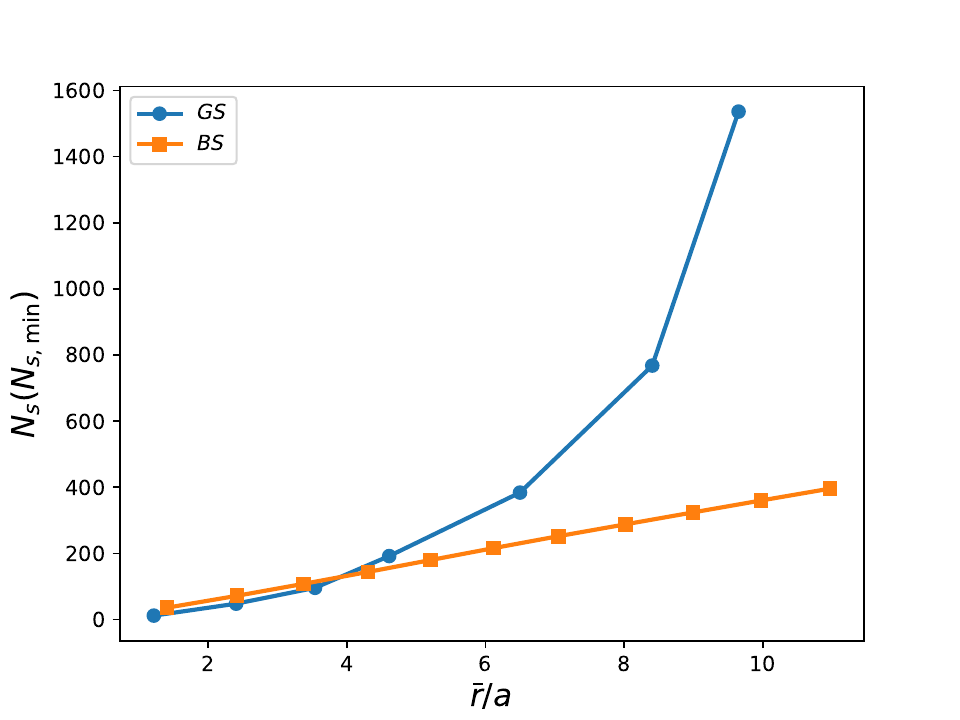}
    \caption{(Minimum required) number of shift operations as a function of smearing size $\bar{r}$ for Gaussian (GS) and the naive block (BS) smearings.}
    \label{fig_gs_r_n}
\end{figure}

Once we have the profile, the smearing size $\bar{r}$ can be defined as the mean-squared-radius of the profile
\begin{equation}
    \bar{r} = \sqrt{\frac{\int \tilde{P}(r)r^2 dr}{\int \tilde{P}(r) dr}}.
\end{equation}
Note that the Gaussian smearing has two parameters, i.e., $\omega$ and $n$.
To understand the relation between $\omega$ and the smearing size $\bar{r}$ with different iteration time $n$,
we carry out similar calculation as in Ref.~\cite{XQCD:2013odc} and the result is shown in Fig.~\ref{fig_gs_w_r_n}. It indicates that for large enough $n$, the smearing size is determined only
by $\omega$ as expected, while when $n$ is not large enough, the smearing size does not grow with increasing $\omega$.
Therefore one can deduce the minimum $n$ for different smearing sizes $\bar{r}$.
The blue points in Fig.~\ref{fig_gs_r_n} show that the minimum time of iterations $n_{\rm min}$ (actually the $y$-axis in the figure is the number of shift operations $N_s=6n_{\rm min}$)
grows exponentially-like as $\bar{r}$ increases. 
In contrast, there is only one parameter $n$ for the naive block smearing,
and, as shown by the orange points in Fig.~\ref{fig_gs_r_n}, the time of shift operations ($N_s=36n$) grows linearly as $\bar{r}$ increases.
As mentioned before, the time cost of a smearing scheme is proportional to the number of shift operations.
Therefore the naive block smearing is much more efficient than the Gaussian smearing in producing large smearing size.

\begin{figure}
    \centering
    \includegraphics[width=0.45\textwidth]{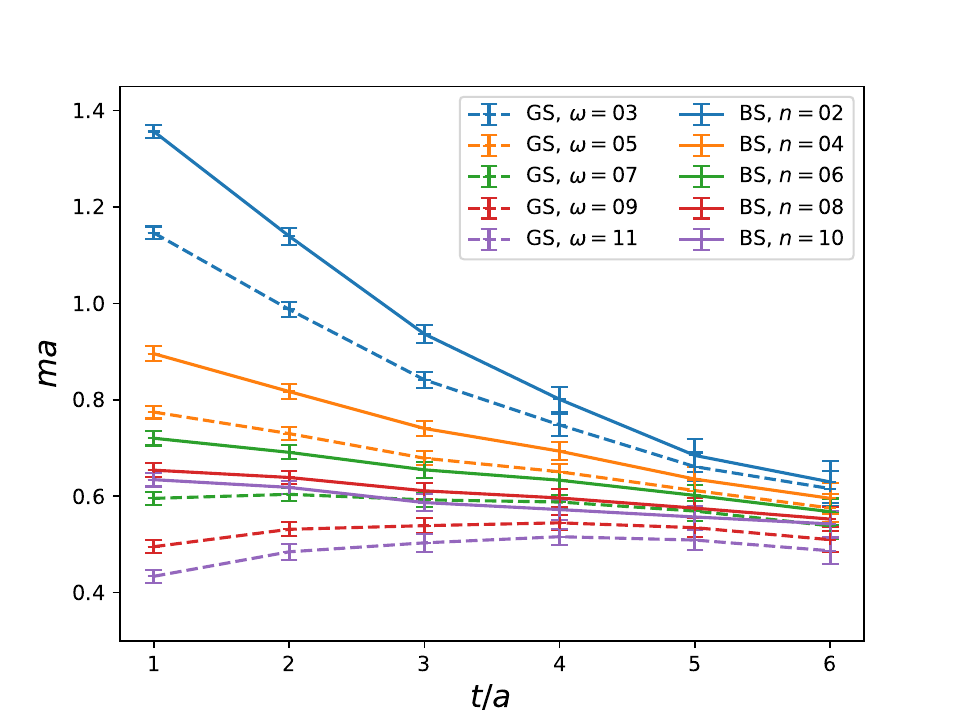}
    \includegraphics[width=0.45\textwidth]{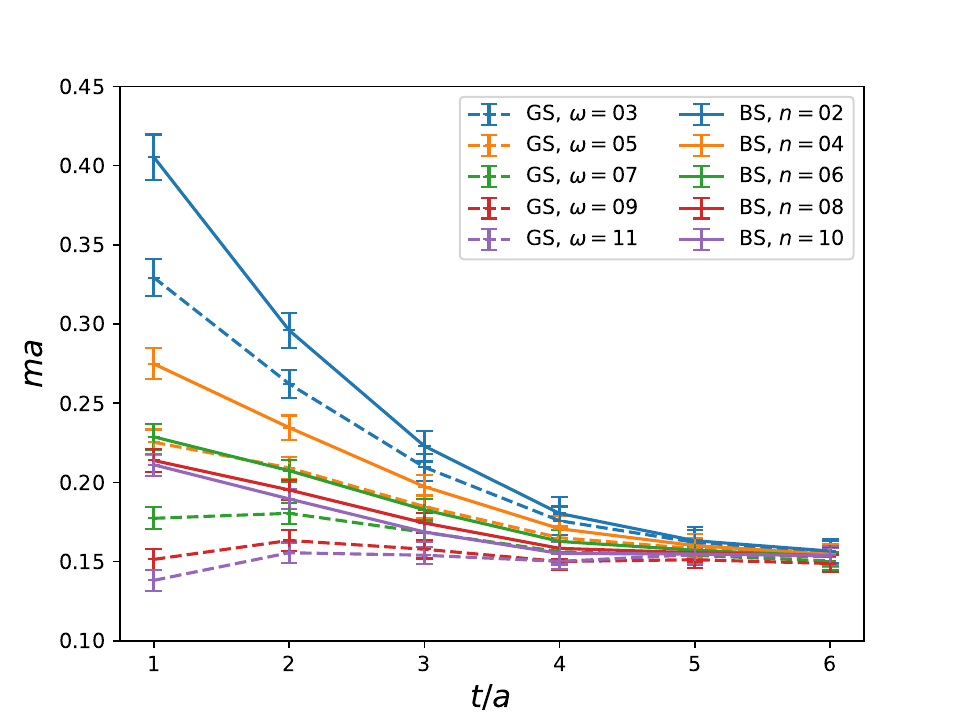} 
    \caption{Pion (left) and proton (right) effective mass for the naive block smearing (BS) and Gaussian smearing (GS) with different smearing parameters.}
     \label{fig_pion_BvG}
\end{figure}

Other than the smearing profile and smearing size, the effects of a smearing scheme can be measured
in practical lattice calculation by looking at the effective mass of hadron two-point correlation functions.
The pion and proton effective mass for the naive block and Gaussian smearings with different smearing parameters are shown in Fig.~\ref{fig_pion_BvG}.
The effective mass plots of the Gaussian smearing show that, with increasing $\omega$,
the excited-state contribution gets smaller and smaller and in the end ($\omega \gtrsim 9$) changes to negative, leading to the bending-up behavior of the effective mass. However, for the naive block smearing, even at $n=$ 8 or 10 where the smearing size $\bar{r}$ is already larger than the Gaussian one with $\omega=$ 11 according to Fig.~\ref{fig_gs_w_r_n} and Fig.~\ref{fig_gs_r_n}, the behavior of the effective
mass is similar to the Gaussian one with $\omega =$ 5 or 7. This indicates that the
smearing profile and size do not reflect all the features of a smearing scheme, especially
the physical effects on the correlation functions,
and the naive block smearing scheme has difficulties in producing strong smearing effects
as far as the effective mass is considered.

\begin{figure}
    \centering
    \includegraphics[width=0.45\textwidth]{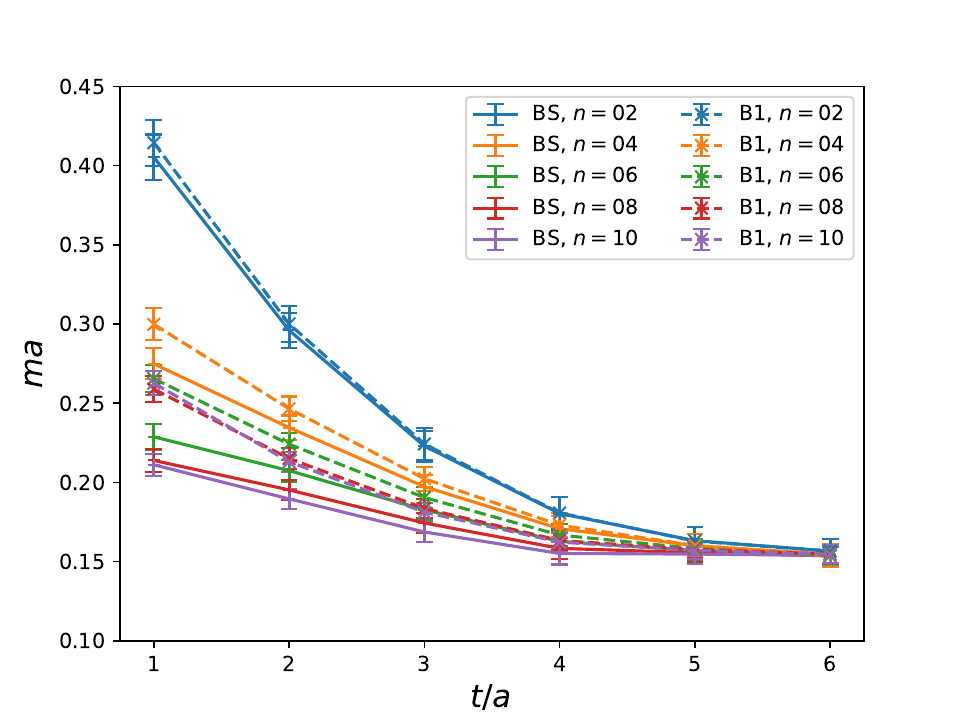}
    \includegraphics[width=0.45\textwidth]{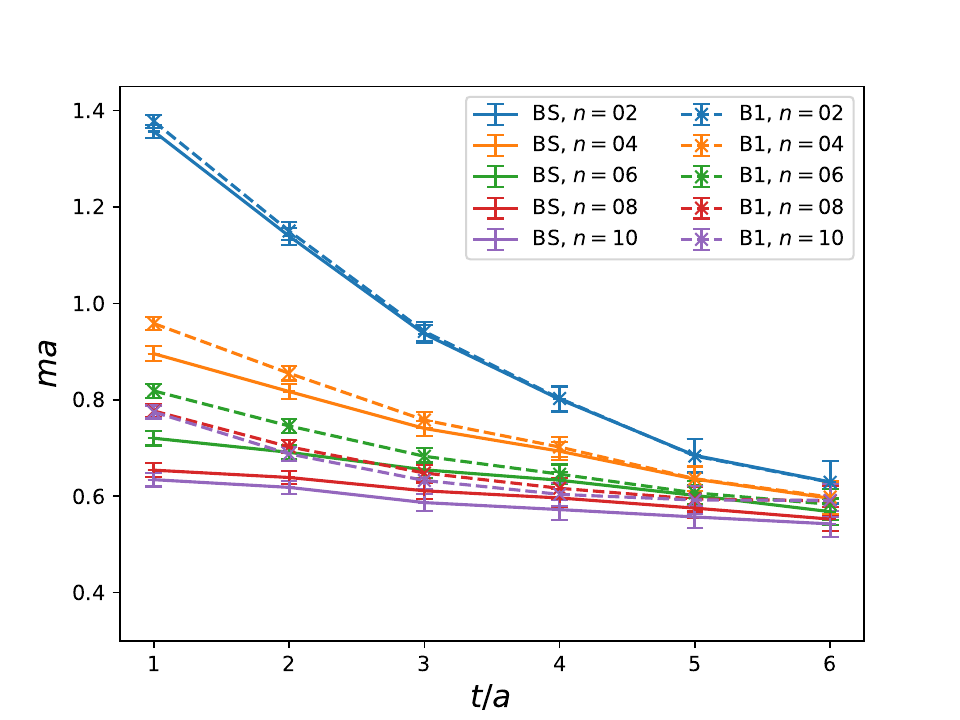}
    \caption{Pion (left) and proton (right) effective masses for the naive block smearing (BS) and the one without summation over the permutations $P(i,j,k)$ in Eq.~(\ref{eq:block}) (B1) with different smearing parameters.}
    \label{fig_pion_BvB1}
\end{figure}

To further investigate the reason, we compare the pion and proton effective masses for the naive block smearing and the one without summation over the permutation $P(i,j,k)$ of Eq.~(\ref{eq:block}) in Fig.~\ref{fig_pion_BvB1}.
For $n=2$, the two results are very close, which implies the potential rotation
symmetry is negligible.
However, for larger $n$, the excited-state contribution of the naive block smearing is better suppressed than that without summation over the permutations.
Recall that the sum over the permutations adds more gauge paths in the smearing function, a natural guess is that the number of gauge links plays an important role in producing strong smearing effects.
Actually, the fact that all possible gauge paths are involved in the Gaussian smearing due to its iterative implementation makes the presumed explanation more reasonable. A novel smearing scheme is proposed in the following section to tackle this problem.

\section{The Novel Smearing Scheme} 
\label{sec:novel_smearing}
Based on the point that the problem of the naive block smearing in producing strong
smearing effects is due to limited number of gauge paths, we construct a new smearing scheme which includes more gauge paths without loss of efficiency.
In this new scheme, instead of summing up the 6 permutations as done
in the naive block smearing, we define the smearing operator as the product of shift operations in different orders of $(i,j,k)$ 
\begin{equation}
    \hat{N} \sim \prod_{(i,j,k)\in P(i,j,k)}^p\left\{
    \left [1 + \sum_{m=1}^{n} \left(({\cal S}^+_i)^m +  ({\cal S}^-_i)^m\right) \right]
    \left [1 + \sum_{m=1}^{n} \left(({\cal S}^+_j)^m +  ({\cal S}^-_j)^m\right) \right]
    \left [1 + \sum_{m=1}^{n} \left(({\cal S}^+_k)^m +  ({\cal S}^-_k)^m\right) \right]\right\},
\end{equation}
where $\prod_{(i,j,k)\in P(i,j,k)}^p$ means we randomly take $p$ $(i,j,k)$ orders from the 6 permutations $P(i,j,k)$, and this 
guarantees that rotation symmetry is restored after configuration averaging.

\begin{figure*}[h]
    \centering    
    \includegraphics[width=0.45\textwidth]{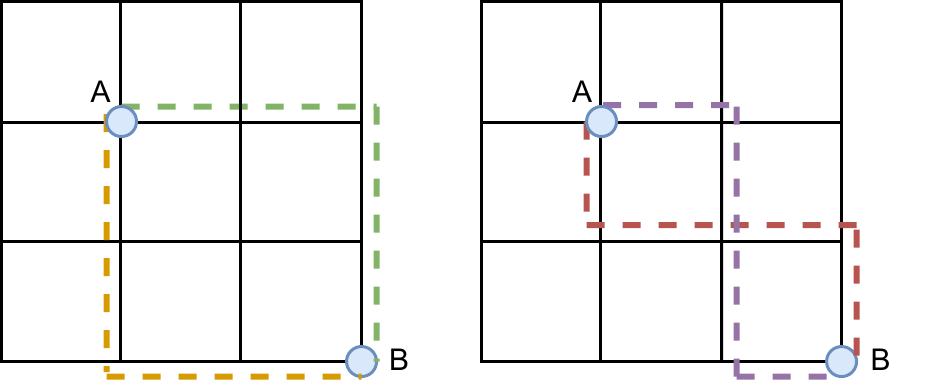}
    \caption{A two-dimensional example of why the novel smearing contains more gauge paths. The dashed lines indicate only two gauge paths connecting point $A$ and $B$ for the naive block smearing (left panel), and the additional shortest paths for the novel smearing (right panel) with $p\ge 2$.}
    \label{fig_NS_path}
\end{figure*}

To understand the path issue, it is key to note that, in the naive block smearing, each gauge path contains at most 3 straight lines of gauge links (i.e., the gauge path can turn at most twice) according to its implementation Eq.~(\ref{eq:block}).
A given order of $(i,j,k)$ gives only one path (e.g. $(i,j,k)=(1, 2, 3)$ means the gauge path going in the $x$ direction first, and then the $y$ direction, and finally $z$), and the summation over the 6 permutations results in 6 paths if there are no degeneration.
A two-dimensional example is shown in the left panel of Fig.~\ref{fig_NS_path}, where in between point $A$ and $B$ there are only two gauge paths denoted as the dashed lines.
Whereas in the novel smearing, shift operations of a specific $(i,j,k)$ order are done on top of the previous shifts, and at most $3p$ straight lines of gauge links (e.g. a possible order of shifts in the $p=2$ case is $x\to y\to z\to y\to z\to x$) can occur.
So there are possibly many more different paths in the novel smearing. 
The additional shortest paths for the novel smearing with $p\ge 2$ are illustrated by the dashed lines on the right panel of Fig.~\ref{fig_NS_path} in the two-dimensional example.

\begin{figure}[h]
    \centering
    \includegraphics[width=0.45\textwidth]{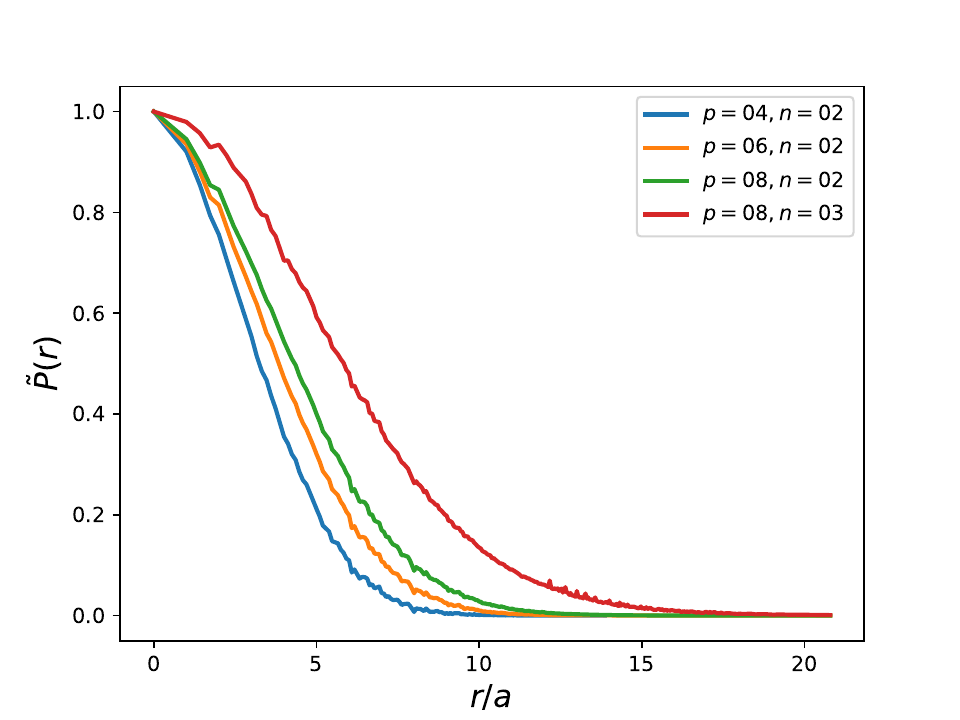}
    \caption{Profile examples of the novel smearing scheme with different $p$ and $n$.}
    \label{fig_prfl_path100}
\end{figure}

\begin{figure*}[h]
    \centering    
    \includegraphics[width=0.85\textwidth]{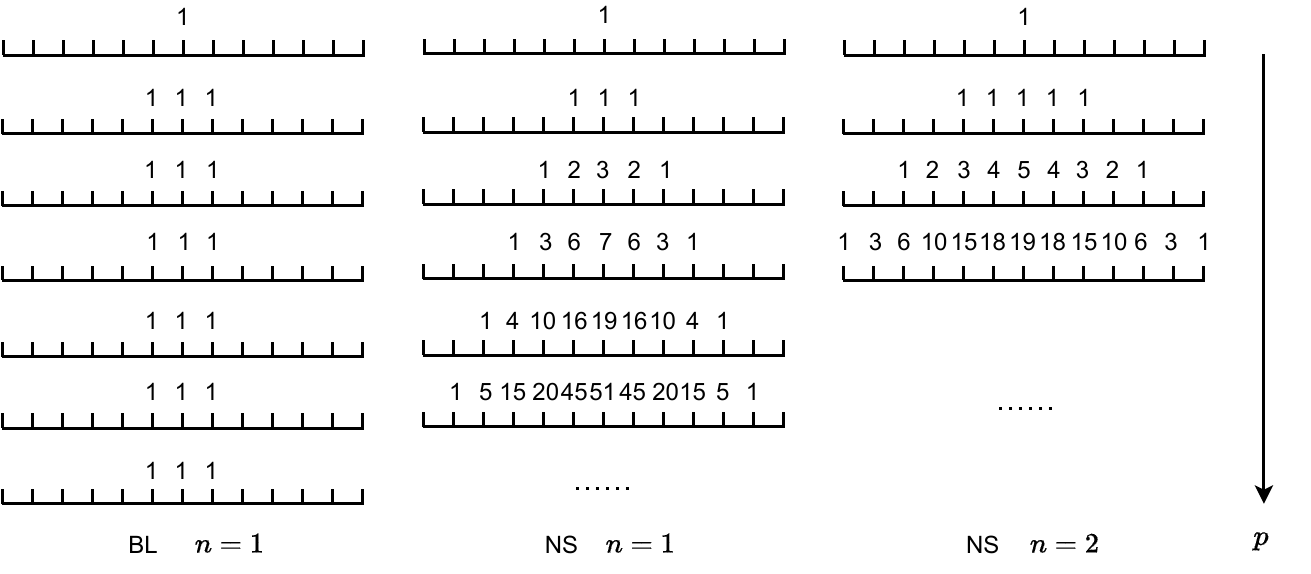}
    \caption{Illustration of the smearing procedure on a one-dimensional unit-gauge lattice.
    The leftmost panel is for the naive block smearing (BS) with $n=1$, and the right two are the profile evolution with increasing $p$ for the novel smearing (NS) with $n=1$ and $2$, respectively.
    }
    \label{fig_NS_cartoon}
\end{figure*}

\begin{figure}
    \centering
    \includegraphics[width=0.45\textwidth]{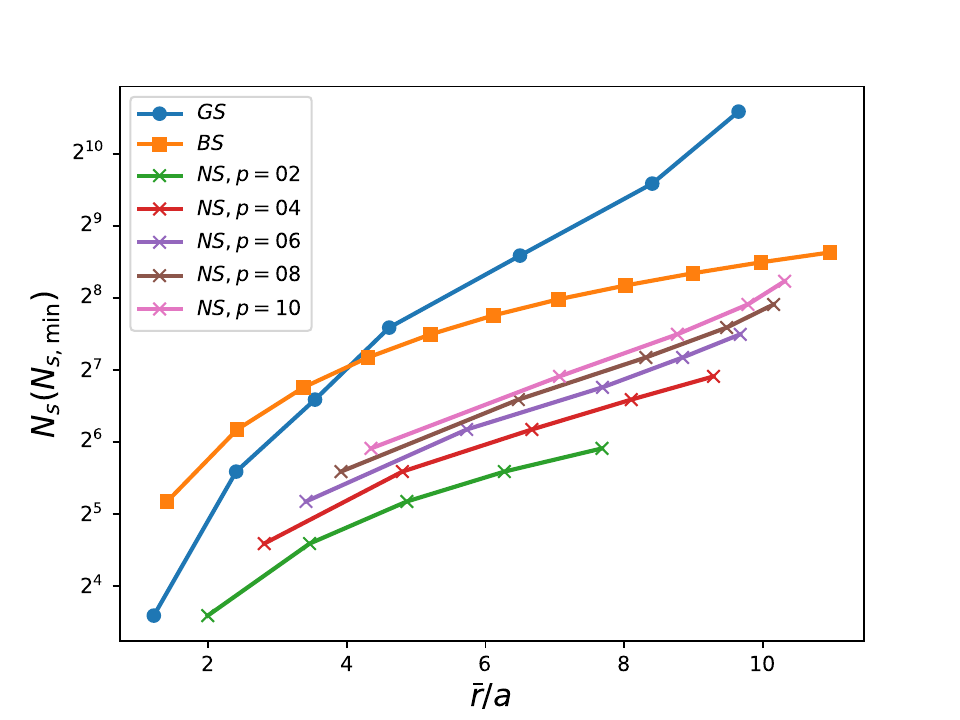}
    \caption{Same as Fig.~\ref{fig_gs_r_n} but including the novel smearing with different $p$.}
    \label{fig_NS_N_r}
\end{figure}

Although the algorithm is modified from the one of the naive block smearing, the profile of the novel smearing is quite different.
As shown in Fig.~\ref{fig_prfl_path100}, the shape of profile looks like Gaussian.
This can be understood intuitively by the illustration of the smearing procedure on a one-dimensional unit-gauge lattice in Fig.~\ref{fig_NS_cartoon}.
The leftmost panel is for the naive block smearing with $n=1$. After smearing, the resultant profile has 3 nonzero points with a constant value. 
The pattern repeats 6 times because the 6 orders of shift operations degenerate for the one-dimensional naive block smearing.
The right two panels represent the novel smearing with $n=1$ and $2$, respectively.
The evolution of the profiles with increasing $p$ is very similar to the Yang Hui's triangle (Pascal's triangle).
The $k_{th}$ number of the $p_{th}$ line of the Yang Hui's triangle equals to the binomial coefficient $C_p^k$ and the corresponding distribution is a binomial distribution with probability $P=1/2$.
The binomial distribution approximates Gaussian distribution very well with large $p$.
And this provides a qualitative explanation why the profile of the novel smearing is Gaussian-like.
A similar plot of the number of shift operations ($N_s=6n\times p$) as a function of smearing size $\bar{r}$ and its comparison with the Gaussian and naive block smearings is shown in Fig.~\ref{fig_NS_N_r}.
We can conclude that the cost of the novel smearing is significantly less than that of the Gaussian smearing for the same $\bar{r}$ (note that the $y$-axis is in log scale),
and for the novel smearing smaller $p$ seems more efficient.
However, as we have emphasized above, the profile and the smearing size are not 
all that matters for a smearing scheme.
The parameter $p$ accounts for the number of paths and small $p$ may not be sufficient to produce strong smearing effects.
Thus, as done in the analysis of the naive block smearing, we examine the smearing effects of the novel smearing with different $p$ and $n$ by calculating pion and proton effective masses.

\begin{figure}
    \centering
    \includegraphics[width=0.45\textwidth]{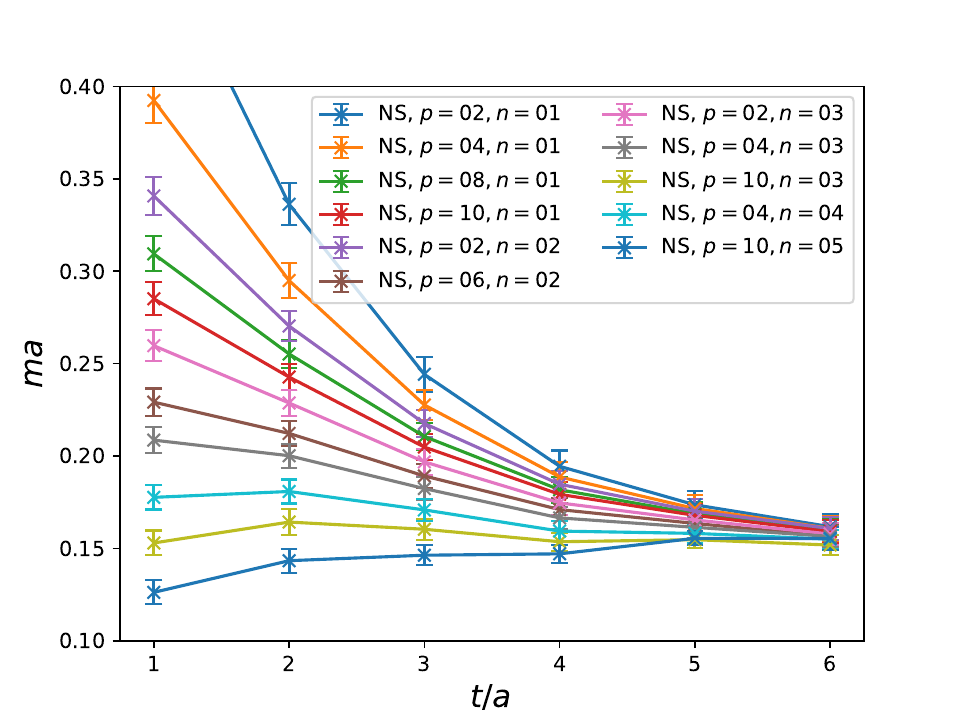}
    \includegraphics[width=0.45\textwidth]{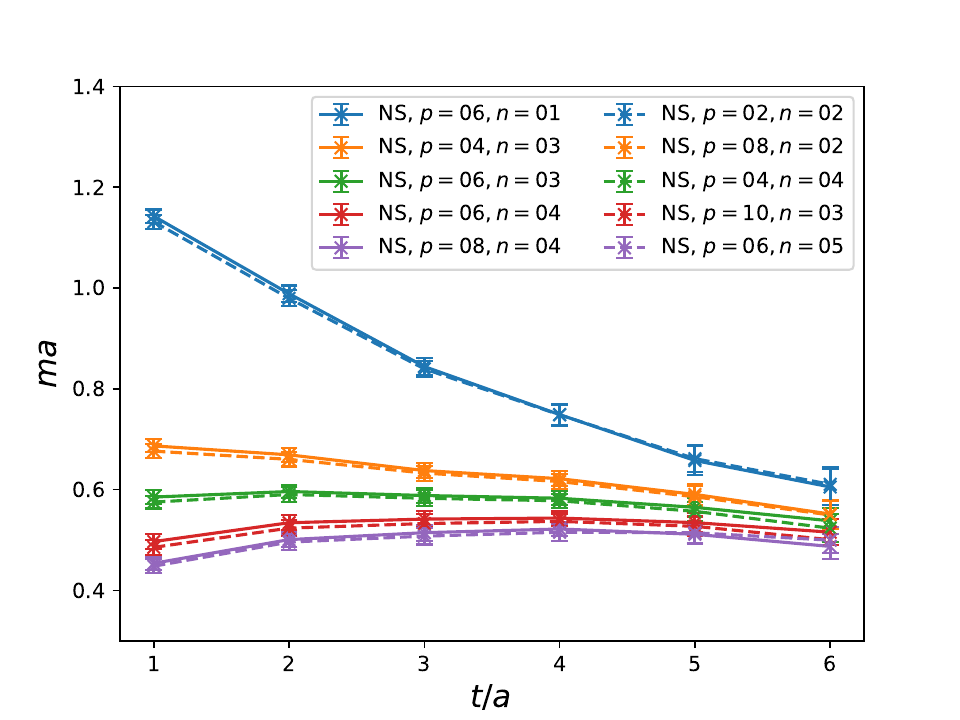}
    \includegraphics[width=0.45\textwidth]{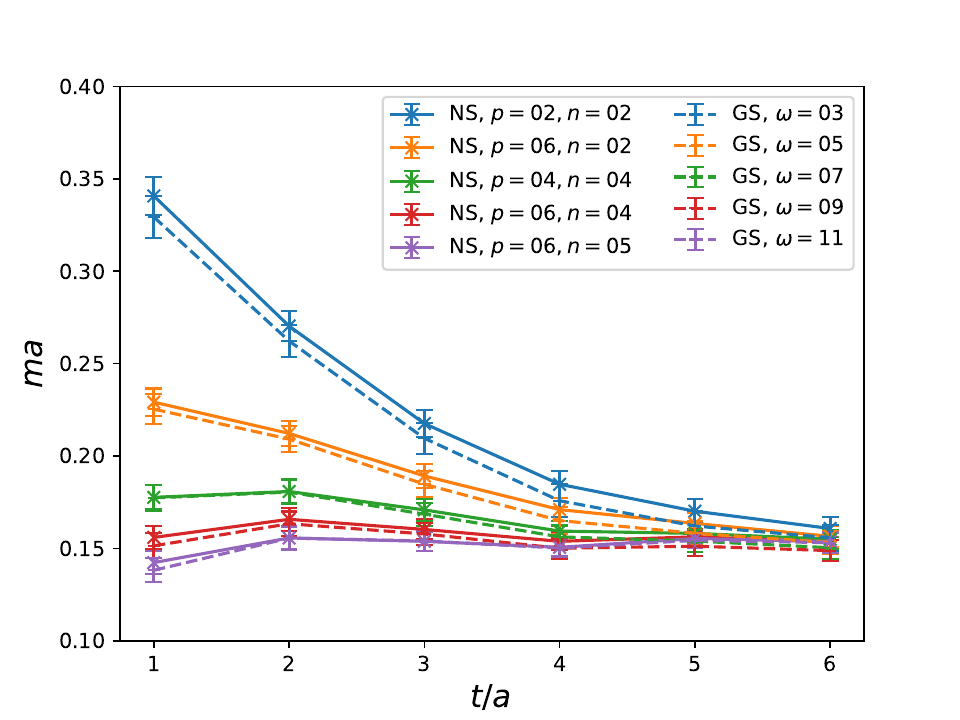}
    \includegraphics[width=0.45\textwidth]{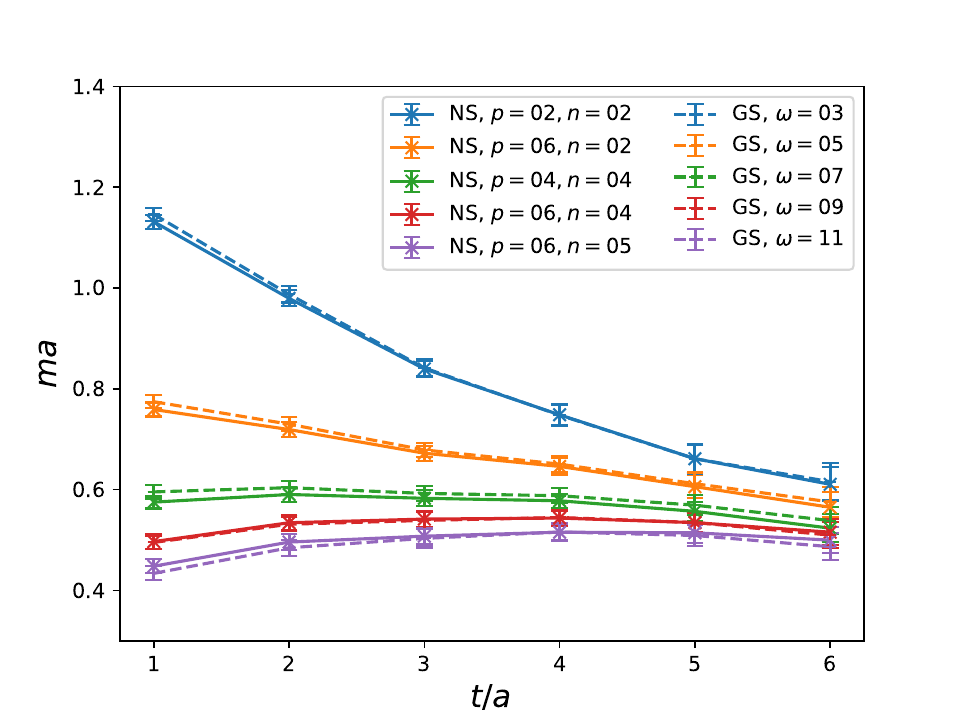}
    \caption{
    The top left panel shows the representative pion effective masses with selected $n$ and $p$ and the top right panel collects some $p$ and $n$ combinations that
    lead to nearly the same proton effective masses.
    Several $n,p$ combinations that give very similar pion and proton effective masses to the Gaussian smearing with typical $\omega$'s are plotted in the lower two panels.
    }
    \label{fig_pion_GSvsNS}
\end{figure}

The effective masses are plotted in Fig.~\ref{fig_pion_GSvsNS}.
The top left panel shows the representative pion effective masses with selected $n$ and $p$, from which we do observe small and even negative excited-state contributions at large $n$ or $p$, indicating that 
this novel smearing scheme can 
resolve the difficulty of the naive block smearing
in producing strong smearing effects caused by the
limited number of gauge links.
The top right panel collects some $p$ and $n$ combinations that
lead to nearly the same proton effective masses.
According to Fig.~\ref{fig_NS_N_r}, 
the combinations with smaller $p$ and bigger $n$ are more effective.
This is qualitatively consistent with our argument that $n$ accounts for enlarging the smearing size while $p$ is mainly responsible for producing more paths.
In this sense, 
when setting the parameters $n$ and $p$ 
for the novel smearing,
one should first change $n$ 
and then use $p$ to fine tune the smearing effects.
We also plot
several
$n,p$ combinations that give very similar pion and proton effective masses to those of the Gaussian smearing
with typical $\omega$'s
in the lower two panels of Fig.~\ref{fig_pion_GSvsNS}.
This gives an empirical way to choose the
parameters in practical calculations,
and we find that for the lattice we use, $p=6$ is large enough
for all different $n$'s.

From the lower panels of Fig.~\ref{fig_pion_GSvsNS}, Fig.~\ref{fig_NS_N_r} and Fig.~\ref{fig_gs_w_r_n}, 
we find that when producing the same smearing effects, 
the cost of the novel smearing is 4 to 8 times improved than the Gaussian smearing,
and the larger the smearing size is the greater the improvement will be.
Specifically, for the Gaussian smearing with $\omega=11$, the corresponding $\bar{r}\sim10a$ and the minimal number of shift operations is around 1536. Whereas for the novel smearing with $n=6$ and $p=5$ which leads to the same effective mass behaviors, only 180 shift operations are needed.
The improvement is about 8 times.
This demonstrates that the novel smearing is an efficient smearing scheme that can be applied in practical lattice studies.

\begin{figure}
    \centering
    \includegraphics[width=0.45\textwidth]{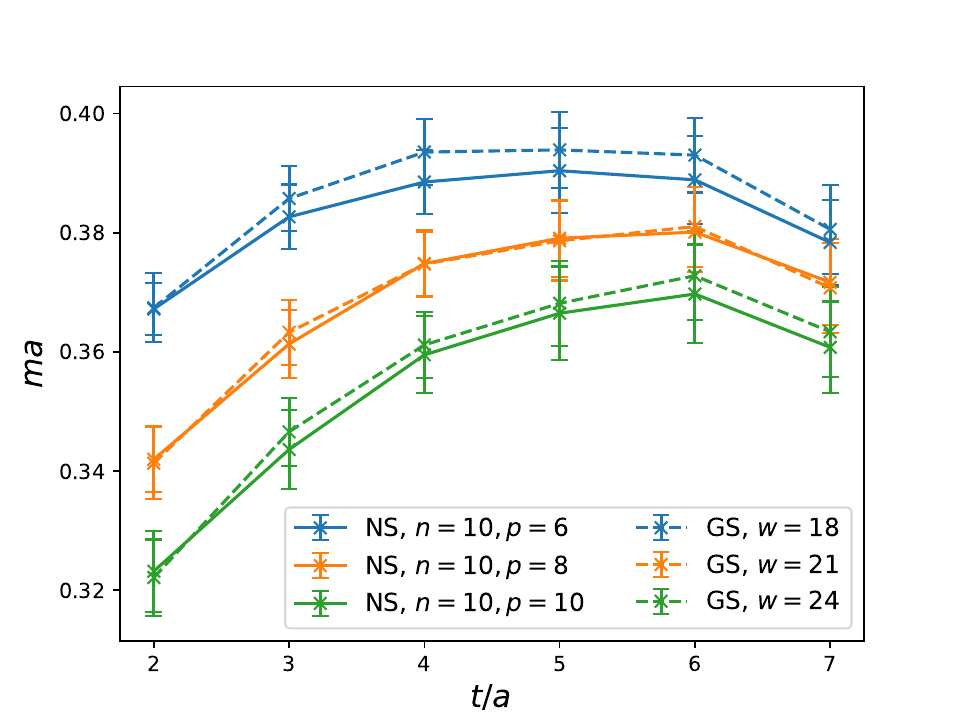}
    \includegraphics[width=0.45\textwidth]{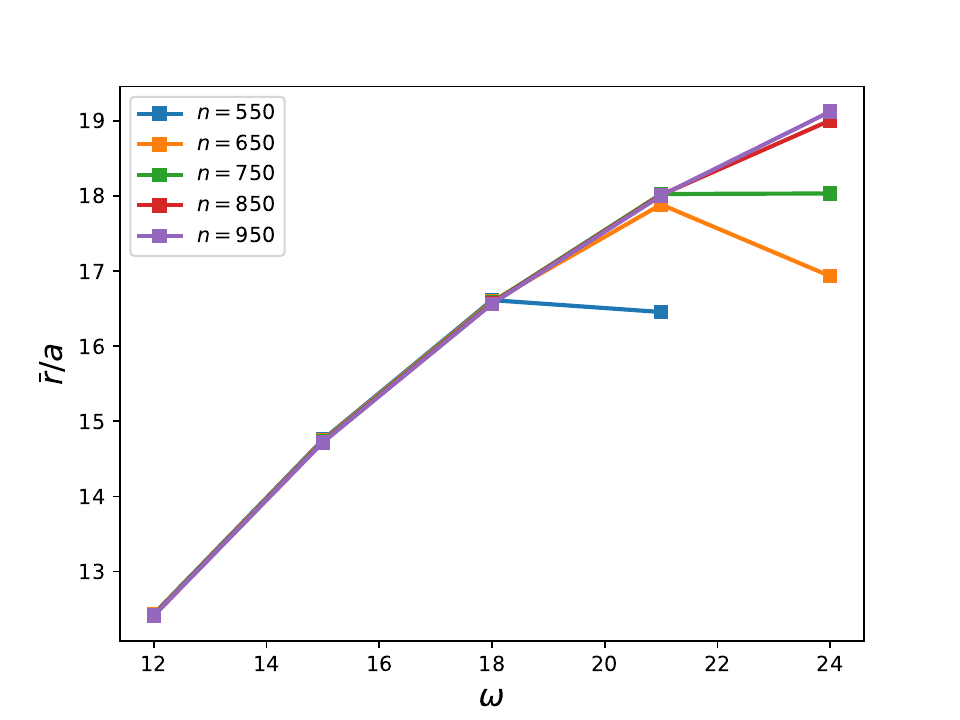}
    \caption{The proton effective masses of the Gaussian smearing and the novel smearing with parameters producing equivalent smearing effects is shown on the left panel.
    The relation between $\omega$ and the smearing size $\bar{r}$ for different iteration time $n$ of the Gaussian smearing is plotted on the right.}
    \label{fig_48pt}
\end{figure}

Actually,
when lattice calculations are
done with finer lattice spacings, heavier quarks or more advanced inverters, the portion of the smearing cost turns larger and becomes compatible with (and even more than) the inversion time cost, and the use of a fast smearing scheme will be more imperative and the improvement will be more substantial.
To give an example, we also carry out the same evaluations on a finer lattice with size of $48^3 \times 144$ and lattice spacing $\sim$ 0.05 fm, which is shown in Fig.~\ref{fig_48pt}.
The Gaussian smearing with $\omega = 24$ and the novel smearing with $n=10$, $p=10$ give
equivalent smearing effects.
The shift operations of the novel smearing and the Gaussian smearing are $6n\times p=600$
and $6\times n_{\text{least}} = 5700$ ($n_{\text{least}}=950$ is read from the right panel of Fig~\ref{fig_48pt}) respectively.
Thus in this case the improvement is around 10 times.

\section{Summary and Outlook}\label{sec:summary}

Based on the naive block smearing, we propose a novel smearing scheme which
gives a Gaussian-like profile and is more efficient than the traditional Gaussian smearing in terms of time cost.
With close smearing effects, 
the improvement of the novel smearing is
about $8 \sim 10$ times compared with the traditional Gaussian smearing for practical smearing sizes $\bar{r}= 10a \sim 20a$.
Besides, by a detailed analysis of the profiles, smearing sizes and hadron effective masses for the naive block smearing, the Gaussian smearing and the novel smearing,
we point out that having a sufficient number of gauge paths is essential for a smearing scheme to produce strong smearing effects.

Although the novel smearing scheme contains
a lot of distinct gauge paths due to the fact that its gauge paths can turn at most $3p-1$ times, the number of gauge paths is still smaller than that for the Gaussian smearing which contains all possible paths.
An important point is that, for $\bar{r}\sim10a$, 
the number of gauge paths in the novel smearing scheme is well enough for $p\sim6$ as indicated by the effective mass behaviors.
Another example is that for small $\bar{r}$, the effective masses of the naive blocking smearing with 6 paths and only 1 path are consistent as shown by Fig.~\ref{fig_pion_BvB1}.
Therefore a qualitative conclusion is that the larger the smearing size is, the more gauge paths we need.
In this sense, the novel path smearing is a fast and effective scheme for practical $\bar{r}$'s used in present lattice studies.
And this scheme can easily accommodate any large smearing sizes  by using a larger parameter $p$.
It is also worth mentioning that our novel smearing is tested to be compatible with the momentum smearing~\cite{Bali:2016lva}.
We believe that this path smearing scheme or its variants should be  inspiring to the community and beneficial for lattice studies of hadron spectra and structures.

\section{Acknowledgement}
This work is partly supported by Guangdong Major Project of Basic and Applied Basic Research under Grant No.\ 2020B0301030008.
JL is supported by the Natural Science Foundation of China (NSFC) under Grants No.\ 12175073 and No.\ 12222503.
JS is supported by
the Natural Science Foundation of China under Grant No.\ 12105108.
TD and KL are supported in part by the Office of Science of the U.S.\ Department of Energy under Grant No. DE-SC0013065 (TD and KL) and No. DE-AC05-06OR23177 (KL), which is within the framework of the TMD Topical Collaboration.
The numerical work has been done on the supercomputing system in the Southern Nuclear Science Computing Center (SNSC).

\bibliography{main}
\end{document}